\documentstyle[preprint,prb,aps,eqsecnum]{revtex}
\begin{document}
\title{
  Spin Gaps and Bilayer Coupling \\
  in YBa$_2$Cu$_3$O$_{7-\delta}$ and YBa$_2$Cu$_4$O$_8$
}
\author{
  A. J. Millis$^{(a)}$ and H. Monien $^{(b)}$\cite{ETH}
}

\address{
  $^{(a)}$ AT\&T Bell Laboratories, 600 Mountain Avenue, Murray Hill,
  NJ 07974\\
  $^{(b)}$ Institute for Theoretical Physics, University of
  California, Santa Barbara, CA 93106
}
\date{\today}
\maketitle
\begin{abstract}
  We investigate the relevance to the physics of underdoped
  YBa$_2$Cu$_3$O$_{\rm 6+x}$ and YBa$_2$Cu$_4$O$_8$ of the quantum critical
  point which occurs in a model of two antiferromagnetically coupled planes of
  antiferromagnetically correlated spins.  We use a Schwinger boson mean field
  theory and a scaling analysis to obtain the phase diagram of the model and
  the temperature and frequency dependence of various susceptibilities and
  relaxation rates.  We distinguish between a low $\omega ,T$ coupled-planes
  regime in which the optic spin excitations are frozen out and a high $\omega
  ,T$ decoupled-planes regime in which the two planes fluctuate independently.
  In the coupled-planes regime the yttrium nuclear relaxation rate at
  low temperatures
  is larger relative to the copper and oxygen rates than would be naively
  expected in a model of uncorrelated planes.  Available data suggest that in
  YBa$_2$Cu$_4$O$_8$ the crossover from the coupled to the decoupled planes
  regime occurs at $T 700K$ or $T \sim 200K$.  The predicted correlation
  length is of order 6
  lattice constants at $T=200K$. Experimental data
  related to the antiferromagnetic susceptibility of YBa$_2$Cu$_4$O$_8$
  may be made consistent
  with the theory, but available data for the uniform susceptibility
  are inconsistent
  with the theory.
\end{abstract}
\narrowtext
\section{Introduction}
In this paper we investigate the relevance to the physics of underdoped
YBa$_2$Cu$_3$O$_{6+x}$ and YBa$_2$Cu$_4$O$_8$ of a $T=0$ order-disorder
transition which occurs in a model of two antiferromagnetically coupled planes
of antiferromagnetically correlated spins.  This transition might be relevant
because: (a) in these compounds the basic structural unit is a pair of CuO$_2$
planes separated from each other by the relatively inert CuO chains
[\onlinecite{structure}], (b) there is evidence for strong antiferromagnetic
correlations within a CuO$_2$ plane [\onlinecite{MMP,Rossad,Tranquada}], (c)
neutron scattering experiments find that for all $\omega$ and $T$ studied a
spin in one plane is perfectly anticorrelated with the nearest neighbor spin
on the nearest-neighbor plane [\onlinecite{Tranquada,Mook}], (d) for $T<150\
{\rm K}$ both the static uniform susceptibility and the various NMR relaxation
rates drop rapidly as $T$ decreases [\onlinecite{Takigawa}] suggesting
[\onlinecite{MM PRL,spingap,Sokol}] that the system is evolving as $T$ is
decreased towards a quantum disordered ground state with a gap to spin
excitations and (e) the spin physics of La$_{\rm 2-x}$Sr$_{\rm x}$CuO$_4$, in
which the CuO$_2$ planes are very weakly coupled, is apparently rather
different [\onlinecite{MM PRL}], suggesting that the behavior of
YBa$_2$Cu$_3$O$_{6+x}$ may be due at least in part to a coupling between
planes.  A preliminary version of this work was published previously
[~\onlinecite{MM PRL}~].

The model we consider is a Heisenberg model of spins sitting on sites of the
lattice depicted in fig. \ref{fig:Lattice} and has two coupling constants,
both taken to be antiferromagnetic. One, $J_1$, couples nearest neighbor spins
in the same plane. The other, $J_2$, couples a spin in one plane to the
nearest spin in the other plane. The Hamiltonian is
\begin{equation}
  H=
  J_1\sum_{<i,j>,a}\vec S^{(a)}_i\cdot\vec S^{(a)}_{j}+
  J_2\sum_i\vec S_i^{(1)}\cdot\vec S_i^{(2)}
  \label{eq:Model}
\end{equation}
Here $i$ labels sites in a given plane, $i$ and $j$ are nearest neighbors in
the same plane, and $a=1,2$ labels two planes.  There are two dimensionless
parameters: $J_2/J_1$ and $S$, the magnitude of the spin.  At temperature
$T=0$ eq.~(\ref{eq:Model}) has two phases.  One is antiferromagnetically
ordered; the other is a singlet phase with a gap to excitations and no long
range order.  Varying $J_2/J_1$ and $S$ produces transitions between the
phases.  In a physical Heisenberg model $S$ would take only half-integer or
integer values $1/2, 1, 3/2, \ldots$\ . However, it is interesting to consider
values $S<1/2$ because the singlet phases occurring for small $S$ in
this model may be useful representations of the ``spin gap'' behavior occurring
in the underdoped YBa$_2$Cu$_3$O$_{\rm 6+x}$.  We shall determine the phase
diagram and discuss the physical properties at the transition and on the
disordered side.  The $T=0$ transition is in the universality class of the 2+1
dimensional Heisenberg model and some universal properties have been
determined [\onlinecite{CHN,Sachdev}].
Here, we pay particular attention to the effects of the interplane coupling,
$J_2$. We show that one must distinguish between a low $\omega ,T$ ``coupled
planes'' regime and a high $\omega ,T$ ``decoupled planes regime''.  In the
coupled-planes regime, one linear combination (essentially the optic mode of
the spin excitation spectrum) is frozen out and the low energy physics is
determined by acoustic spin fluctuations in which moments in the two planes
fluctuate coherently.  In the decoupled planes regime the two planes fluctuate
essentially independently. One important feature of the quantum critical point
considered here is the close relationship between the susceptibility at small
$q$ and at a $q$ near the ordering wavevector [\onlinecite{Sachdev}].  We show
that this, when combined with interplanar coupling, has a surprising
implication for the yttrium relaxation rate: in the coupled-planes regime it
is larger, relative to the other rates, than one would expect from a model of
uncoupled planes.

We emphasize that eq.~(\ref{eq:Model}) is not a completely realistic model of
YBa$_2$Cu$_3$O$_{6.6}$ or YBa$_2$Cu$_4$O$_8$ because it omits the itinerant
carriers which make these materials metallic and indeed superconducting. The
itinerant carriers also strongly affect the magnetism. In all of the
hole-doped CuO$_2$ compounds, long-range magnetic order disappears essentially
at the metal-insulator transition [\onlinecite{LRO}]. It therefore seems
likely that the itinerant carriers substantially weaken the in-plane magnetic
correlations.  Itinerant carriers also presumably give rise to a particle-hole
continuum of incoherent spin excitations which may strongly affect the physics
and which in appropriate circumstances may change the universality class of
the transition [\onlinecite{Hertz,AJM}].  It has however recently been argued
that there is some evidence that the behavior of YBa$_2$Cu$_3$O$_{6.6}$ and
YBa$_2$Cu$_4$O$_8$ is in the universality class of eq.~(\ref{eq:Model})
[\onlinecite{Sokol}].

The rest of the paper is organized as follows. In section II we solve
eq.~(\ref{eq:Model}) via the Schwinger-boson mean field theory. In section III
we give the results of the Schwinger boson mean field theory for the phase
diagram and the physical susceptibilities relevant to NMR and neutron
scattering experiments.  In section IV we combine selected results of the mean
field theory with other arguments to map the problem onto a recently
constructed scaling theory of the transition [\onlinecite{Sachdev}] and give
the relevant results of the scaling analysis.  Section V is a conclusion in
which we discuss the results and their relation to experiments on
YBa$_2$Cu$_3$O$_{6.6}$ and YBa$_2$Cu$_4$O$_8$. It may be read independently of
the previous sections by readers uninterested in the derivations of the
results. Appendices present details of various calculations.

\section{Mean Field Solution}
\label{sec:Mean Field}

We first study eq.~(\ref{eq:Model}) by the Schwinger-boson mean-field method
[\onlinecite{Arovas Auerbach}]. In this method one introduces bose operators
$b^{\dagger(a)}_{i\alpha}$ which create a state of spin $\alpha$ on site $i$
of plane a.  One restricts oneself to the subspace in which each site on each
plane has $2S$ bosons, corresponding to spin $S$; thus we enforce the
constraint
\begin{equation}
  \sum_\alpha b^{\dagger(a)}_{i\alpha}b^{(a)}_{i\alpha}= 1 + 2S
  \label{eq:Constraint}
\end{equation}
A spin operator is written
\begin{equation}
  \vec S_i^{(a)}=
  \sum_{\alpha\beta}b^{\dagger(a)}_{i\alpha}\vec\sigma_{\alpha\beta}
  b^{\phantom{\dagger}(a)}_{i\beta}
  \label{eq:Spin}
\end{equation}
We now substitute eq.~(\ref{eq:Spin}) into eq.~(\ref{eq:Model}).  Then on the
even sublattice of plane 1 and the odd sublattice of plane 2 we make the time
reversal transformation which in the $S=1/2$ case is:
\begin{eqnarray}
  b^\dagger_\uparrow&\to&-b^\dagger_\downarrow\cr
  b^\dagger_\downarrow&\to&\phantom{-}b^\dagger_\uparrow
  \label{eq:Staggering}
\end{eqnarray}
Rearranging and using eq.~(\ref{eq:Constraint}) leads to
\begin{eqnarray}
  H'&=&-{1\over 2}J_1\sum_{<i,ja\alpha\beta}\Bigg(b^{\dagger(a)}_{i,\alpha}
  b^{\dagger(a)}_{j,\alpha}\Bigg)
\Bigg(b^{(a)}_{i\beta}b^{(a)}_{j,\beta}\Bigg)\cr
  &-&{1\over2}J_2\sum_{i\alpha\beta}
  \Bigg(b^{\dagger(1)}_{i,\alpha}b^{\dagger(2)}_{i,\alpha}\Bigg)
  \Bigg(b^{(1)}_{i,\beta}b^{(2)}_{i,\beta}\Bigg)
  \label{eq:HBoson}
\end{eqnarray}

To proceed with the approximate mean-field treatment one introduces a Lagrange
multiplier $\mu$ to enforce the constraint, and an in-plane bond field $Q$ and
a between-planes bond-field $\Delta$ to decouple the quartic interactions in
eq.~(\ref{eq:Model}). In the mean field approximation $\mu$, $\Delta$ and $Q$
are taken to be constant in space and time.  The resulting theory may be
diagonalized.  The manipulations are standard and are given in Appendix A.
The result is a model of two species of bosons ($s$, for symmetric under
interchange of planes and $a$, for antisymmetric under interchange of planes)
governed by the Lagrangian
\begin{equation}
  {\cal L}'_B =
  \sum_{k\alpha}
  s^\dagger_{k\alpha}[\partial_\tau+\omega_k]s_{k\alpha}+a^\dagger_{k\alpha}
  [\partial_\tau+\omega_{k+P}]a_{k\alpha}
  \label{eq:LB'}
\end{equation}
with
\begin{equation}
  \omega_k=\sqrt{\mu^2-(Q\gamma_k+\Delta)^2} ,
\end{equation}
\begin{equation}
  \gamma_k={1\over2}(\cos k_x+\cos k_y)
  \label{eq:A-Dispersion}
\end{equation}
and
\begin{equation}
  P=(\pi,\pi).
\end{equation}
Note that $\gamma_{k+P}=-\gamma_k$. The parameters $\mu$, $Q$ and $\Delta$ are
determined by the mean-field equations
\begin{mathletters}
  \label{eq:MF}
  \begin{eqnarray}
    \int\frac{d^2k}{(2\pi)^2} {\mu\over\omega_k}\coth{\omega_k\over2T}&=& 1 +
2S,\\
    \int\frac{d^2k}{(2\pi)^2} {(Q\gamma_k+\Delta)\gamma_k\over\omega_k}\coth
    {\omega_k\over2T}&=&Q/ 2 J_1, \\
    \int\frac{d^2k}{(2\pi)^2}
    {Q\gamma_k+\Delta\over\omega_k}\coth{\omega_k\over2T}&=&2 \Delta/J_2.
  \end{eqnarray}
\end{mathletters}
These equations are derived in Appendix A and imply $Q,\Delta,\mu$ are real,
and $\Delta,\mu>0$.  Further, one may change the sign of $Q$ by shifting the
origin of reciprocal space to $k=P$ and interchanging the labels $s$ and $a$.
Thus we take $Q>0$ with no loss of generality.

The integrals in eqs.~(\ref{eq:MF}) depend on $k$ only via $\gamma_k$; one may
therefore recast them as $\int{d^2k\over(2\pi)^2}\to\int d\gamma N(\gamma)$
where $N(\gamma)$ is a density of states which is constant near the band edges
$\gamma=\pm1$ and logarithmically divergent at the band center $\gamma=0$.  To
obtain an analytically tractable model we replace $N(\gamma)$ by $1/2$. The
resulting equations are solved in Appendix B.  At $T=0$ we find the phase
diagram shown in fig. 2.  At $J_2=0$ we have two decoupled planes.  As is well
known [\onlinecite{CHN,Sachdev,Arovas Auerbach,Sachdev and Read}] a single
plane has a transition at $S=S_c\cong(\pi /2-1)/2\cong 0.28$ (the numerical
expression for $S_c$ is obtained from the mean-field calculation) between a
large-$S$ ordered state and a small-$S$ singlet state with a gap to all spin
excitations.  If one increases $J_2$ in the ordered ($S>S_c$) phase of the
one-plane model, one reaches at a $J_2 \sim\; J_1$ another transition line, at
which the ordered phase is destroyed in favor of singlets which are
principally between planes.  At $S=1/2$ the Schwinger boson mean field method
yields a second order transition at $J_2\cong4.48\;J_1$.  A previous series
expansion study of eq.~(\ref{eq:Model}) by Hida [\onlinecite{Hida}] yielded a
second-order transition at $J_2\cong2.56\;J_1$ and a very recent Monte-Carlo
study by Sandvik [\onlinecite{Sandvik}] found a second order transition at
$J_2 \cong (2.7 \pm 0.2 )\;J_1$.

Interestingly, for $S<S_c$, we find that increasing $J_2$ from $J_2=0$
initially reduces the gap in the singlet phase, thus moving the system closer
to order.  For $S^*<S<S_c$ (with $S^*\cong0.19$ in the mean field calculation)
the phase diagram is reentrant.  We believe that the physics behind the
reentrance is that a small $J_2$ splits the spectrum into acoustic and optic
sectors, and because the optic sector involves a coupled motion of the spins
in the two planes the effective spin of the model describing the low energy
fluctuations is increased, promoting order, whereas at large $J_2$ the
between-planes interaction produced singlets, favoring destruction of order.

In the disordered phase, $\mu>(Q+\Delta)$ and the excitation spectrum has a
gap at all wavevectors.  Two gaps that will be particularly important in what
follows are $\omega_+$ and $\omega_-$, given by
\begin{equation}
  \omega_+ = \sqrt{\mu^2 -(Q+\Delta)^2}
  \label{eq:Defomega+}
\end{equation}
and
\begin{equation}
  \omega_- = \sqrt{\mu^2 -(Q-\Delta)^2}.
  \label{eq:Defomega-}
\end{equation}
For the $s$-bosons $\omega_+$ is the gap at $k=0$ and $\omega_-$ is the gap at
$k=P$; for the $a$-bosons 0 and P are interchanged. We shall see that the
matrix elements coupling the bosons to externally applied fields have a strong
$k$-dependence, so that measurable susceptibilities near $k=0$ differ
dramatically from those near $k=P$. Both gap parameters are temperature
dependent.

At the $T=0$ phase transition, $\omega_+$ vanishes while $\omega_- >0$. Near
the large $J_2$ boundary of the ordered phase, we have $\Delta \sim Q$ and
$\omega_- \sim J_1 \gg \omega_+$.  The $s$-boson mode has one low energy
branch, centered at $k=0$,
\begin{equation}
  \omega_+(k)^2 = \omega_+^2+2\mu^2(1-\gamma_k) = \omega_+^2+v^2k^2
  \label{eq:omega+}
\end{equation}
Here we have expanded $\gamma_k = 1-k^2/4$, set $Q+\Delta = \mu$ and defined
$v^2 = \mu^2/2$.

In the lower, reentrant branch of the phase diagram, i.e. at $T=0$,
$S^*<S<S_c$, $J_2 \ll J_1$ we find from eqs.~(\ref{eq:SmallJ2}) that the phase
boundary is given by
\begin{equation}
  \frac {J^*_2} {J_1} = \pi (S_c-S) + (\pi^2-8)(S-S_c)^2
\end{equation}
On the phase boundary,
\begin{equation}
  \omega_- = J_2 + ...
\end{equation}
Here the ellipsis denotes terms of order $(S_c-S)^3$, $(J_2/J_1)^3$ and higher.
If we tune through the phase transition by varying $J_2$ we find that
sufficiently deep in the ordered phase $\omega_-$ increases as $(J_2J_1)^{1/2}$
as expected from spin-wave theory [\onlinecite{Tranquada}], while $\omega_-$
approaches $\omega_+$ very rapidly as $J_2$ is decreased into the disordered
phase.  Indeed, within mean field theory we find find that for
$J_2/J_1 \le \pi (S_c-S)-(8-\pi^2/2)(S_c-S)^2$ a solution with
$\omega_+ \ne \omega_-$ is not possible.  The sharp transition from a solution
with $\omega_- \omega_+$ to  one with $\omega_- = \omega_+$ is an artifact
of mean field theory, but the qualitative result, that
$(\omega_--\omega_+)/\omega_-$ drops rapidly as one moves into the disordered
phase, is likely to be correct for this model.  The argument is that the ground
state of the one plane model in the disordered phase is a singlet with a gap
of order $J_1(S_c-S)$; for $J_2$ less than this value the interplane coupling
can only slightly perturb the singlets.

For $S=S_c$ and $J_2 \ll J_1$ the $s$-boson has two low energy branches, one
centered at $k=0$ with dispersion given by eq. (\ref{eq:omega+}) and one given
by
\begin{equation}
  \omega_-(k)^2 = \omega_-^2+2\mu^2(1+\gamma_k)= \omega_-^2+v^2(k-P)^2
  \label{eq:omega-}
\end{equation}
Here $v=\mu/2 +...$ and the ellipsis indicates terms of order $(J_2/J_1)^2$.

We now consider $T>0$.  In a realistic model there are no phase transitions,
however different regimes of behavior exist.  In the mean field theory,
crossovers between different regimes sometimes appear as unphysical phase
transitions.  We are interested in properties in the disordered regime near
the critical line.  At the $T=0$ phase transition, $\omega_+$ vanishes; close
to it $\omega_+$ is much smaller than $J_1$.  For $T < \omega_+(T=0)$ the
number of thermal excitations is negligible; the physics is of a singlet
ground-state with a $q$-dependent gap to excitations.  In the literature this
is referred to as a "quantum disordered" regime.  In this regime the low
energy spectrum only involves one linear combination of the spin excitations
in the two different planes; the antisymmetric one for $k$ near $P$ and the
symmetric one for $k$ near 0.  For $\omega_+(T=0)<T<\omega_-(T=0)$ this one
linear combination becomes thermally excited.  We refer to this as the
"coupled-planes quantum critical regime".  It only exists if
$\omega_-(T=0)-\omega_+(T=0)$ is large enough.  In the coupled-planes critical
regime $\omega_+(T) \sim T$, but $\omega_-(T)$ takes its zero temperature
value.  The crossover from the quantum disordered regime to this quantum
critical regime is identical to that occurring in a one-plane model.  Finally,
as $T$ is increased through $\omega_-(T=0)$ the other linear combination of
spin excitations also becomes excited and the two planes begin to fluctuate
more or less independently.  If $\omega_-(T=0) \ll J_1$ then we find in the
mean field theory that the behavior in this regime will be controlled by the
$T=0$ critical point of a single plane.  We refer to this regime as the
"decoupled-planes critical regime".  In the mean field theory the change from
the coupled-planes to the decoupled-planes regime occurs via a second order
phase transition.  We expect fluctuations not included in the mean field
theory will convert this into a smooth crossover.  In the decoupled-planes
regime, both $\omega_+$ and $\omega_-$ are proportional to $T$ with the same
coefficient.  Fig. 3 shows the $T$-dependence of $\omega_+$ and $\omega_-$ for
parameters chosen so that $\omega_+(T=0)=0$, and indicates the different
regimes.
\section{Physical Quantities}
\label{sec:Physical}

To obtain the magnetic susceptibilities we compute the linear response of the
system to an externally applied magnetic field $\vec h_i^{(a)}$.  The details
are given in Appendix C.
We find it convenient to decompose the externally applied field into parts
symmetric and antisymmetric under interchange of planes,
and compute the linear response to
\begin{equation}
  \Delta H
  =\sum_q {h^{(1)}_q-h^{(2)}_q\over2}O^a_q+{h^{(1)}_q+h^{(2)}_q\over2}O^s_q
\end{equation}
Here $O^s$ and $O^a$  are operators creating spin fluctuations symmetric and
antisymmetric under interchange of planes respectively.
The only non-zero susceptibilities are
\begin{mathletters}
  \label{eq:chi}
  \begin{eqnarray}
    \chi^{aa}_q(\omega)
    &=&\int^\infty_0 dt e^{i(\omega+i\epsilon)t}
    <{[O^a_q(t),O^a_{-q}(0)]}>
    \label{eq:chiaa}
  \end{eqnarray}
  and
  \begin{eqnarray}
    \chi^{ss}_q(\omega)
    &=&\int^\infty_0 dt e^{i(\omega+i\epsilon)t}
    <{[O^S_q(t),O^S_{-q}(0)]}>
    \label{eq:chiss}
  \end{eqnarray}
\end{mathletters}

The uniform susceptibility (written as a susceptibility per spin
and \underline{not} as a susceptibility per unit cell) is
\begin{equation}
  \chi(T)=\lim_{q \to 0} \chi^{ss}_q(\omega=0)={1\over4T}\sum_k\sinh^{-2}
  \left(\frac{\omega_k}{2T}\right)
  \label{eq:chi0}
\end{equation}
We note that in the approximation of section II and the Appendices, namely,
\begin{equation}
  {kdk\over2\pi}=N(\gamma)d\gamma={\omega d\omega \over 2}
  \label{eq:k->omega}
\end{equation}
eq.~(\ref{eq:chi0}) becomes
\begin{equation}
  \chi(T)= 2T [
  \int^\infty_{\omega_+/2T} \frac {dxx} {\sinh^2(x)} +
  \int^\infty_{\omega_-/2T} \frac {dxx} {\sinh^2(x)}
  ]
  \label{eq:chiu}
\end{equation}
 From eq.~(\ref{eq:chiu}) we see that in the coupled-plane quantum disordered
regime $(\omega_+(T=0) \gg T)$, $\chi (T) \sim \omega_+e^{-\omega_+/T}$; this
is a factor of two smaller than would be found for a single plane in the
disordered regime
with the same gap.  In the coupled-plane critical regime
$\omega_+(T=0) \ll T \ll \omega_-(T=0)$, $\chi (T) \sim T$ and in the
decoupled-plane critical regime $T \ll \omega_-(T=0)$, we also have
$\chi (T) \sim T$ but with a coefficient larger by a factor of two.
This behavior is easy to understand:  in the coupled-plane
regime one of the two spin degrees of freedom per unit cell is frozen out; in
the decoupled-plane region both are free to fluctuate.

Next we consider $\chi^{''}(q,\omega)$ for $q=P$.
 From the results of eq. (\ref{eq:chiaa}, \ref{eq:chiss})
we find, for $\omega 0$

\begin{equation}
  \chi_P^{aa''}(\omega ,T) = \frac{\pi} {\omega} \coth(\omega / 4T)
  \left[
  \Theta (\omega-2\omega_-)+\Theta(\omega-2\omega_+)
  \right]
\label{eq:chiaaP}
\end{equation}

\begin{eqnarray}
  \chi_P^{ss''}(\omega ,T) &=& \frac{2 \pi } {\omega} \big[
  \Theta (\omega-(\omega_++ \omega_-))
  [
  1 + b((\omega^2-\omega_+^2+\omega_-^2)/2 \omega) +
  b((\omega^2+\omega_+^2-\omega_-^2)/2 \omega)
  ] \cr
  &+&
  \Theta(\omega_--\omega_+-\omega)
  [
  b((\omega_-^2-\omega_+^2-\omega^2)/2 \omega) -
  b((\omega_-^2-\omega_+^2+\omega^2)/2 \omega)
  ]
  \big]
  \label{eq:chissP}
\end{eqnarray}

These formulae are plotted in fig. 5 for the parameters used to construct fig.
3. The sharp onset at $T=0$ is an unphysical feature of the dynamics in the
mean field theory, which is removed when fluctuations are included
[\onlinecite{Sachdev,Chubukov}].  A related peculiarity of the mean field
theory is that $\chi'(q,\omega =0)$ decays as $1/q$ for $q$ near $P$.  We
believe that despite the obvious artificialities the mean field expressions
give some reliable information about the spin excitations of the model, as in
the case of the one-plane Heisenberg model [\onlinecite{Arovas Auerbach}].  In
particular, we believe that the location of the peaks and their relative
magnitudes and temperature dependences give a reasonable representation of the
location, relative magnitude and temperature dependence of the peaks in the
appropriate susceptiblities of the model.  We see that from eqs.
(\ref{eq:chiaaP}) and (\ref{eq:chissP}) and fig. 5 that at low $T$ the
important energy scale for $\chi_P^{aa''}$ is $2 \omega_+$ while for
$\chi_P^{ss''}$ it is $\omega_++\omega_-$.  Everywhere in the disordered phase
the difference between these energies is less than $J_2$.  It is also clear
that until the temperature becomes comparable to the scale $T^*$ at which the
crossover from the coupled-planes to the decoupled-planes critical regime
occurs, $\chi_P^{aa''} \gg \chi_P^{ss''}$, and that $\chi_P^{ss''}$ is
confined mostly to high frequencies, of order $2J_2$.  For this reason we
suspect that the neutrons scattering data, which have not observed any
$\chi_P^{ss''}$ component for $\omega < 40\ {\rm meV}$, do not set a stringent
limit on $J_2$, although they suggest that it is greater than $20\ {\rm meV}$.

We now turn to the relaxation rates.  Details of the computations may be found
in  Appendix C.  We begin with that of Yttrium, $1/{^YT_1T}$, which is found
to be
\begin{equation}
  \frac{1}{^YT_1T} = 2 \pi D^2  \lim_{\omega\to 0}\frac{1}{\omega} \sum_q\
|g(q)|^2
  \chi^{ss''}_q(\omega)
\end{equation}
The form factor $g$ is equal to 4 for $q$ near 0 and to $(q-P)^2$ for $q$ near
$P$; a more precise formula is given in eq. (\ref{eq:C-1/YT1T}).

Using eqs.~(\ref{eq:k->omega}) and (\ref{eq:C-1/YT1T}) we find
\begin{equation}
  \frac{1}{{^YT}_1T}=  8\pi D^2 T^2
  \left[
  \int^\infty_{{\omega_+(T)/2T}} \frac{dxx^2} {sinh^2(x)}
  +
  \int^\infty_{{\omega_-(T)/2T}} \frac{dxx^2} {sinh^2(x)}
\right]
\label{eq:yT1}
\end{equation}
In the mean field approximation to both the coupled planes and the decoupled
planes critical regimes, the yttrium rate $1/{{^YT}_1T} \sim T^2$, so it
vanishes faster than the static susceptibility, while in the low $T$ limit
$1/{{^YT}_1T} \sim e^{-\omega_+/T}$ is proportional to the uniform
susceptibility.  Note that the ratio of $1/^YT_1T$ to $\chi^2$ is the same in
the coupled-plane critical regime as it is in the decoupled-plane critical
regime.  There are two compensating effects at work: in the coupled-plane
regime there is only one spin mode at small $q$ (instead of the two that occur
in a model of two uncoupled planes) but in this mode the 8 spins add
coherently to the yttrium relaxation rate, whereas in the decoupled-plane
regime there are twice as many spin excitations (so one would naively expect
the rate to be four times as large) but the two planes add incoherently
(reducing the rate by a factor of two). $1/^YT_1T$ is plotted in fig. 4 for
the parameters used to construct fig. 3.

We now consider the oxygen rate.  This has one contribution from the small $q$
fluctuations and one from fluctuations with $q$ near $P$.  The latter is
suppressed by a form factor because each oxygen sits symmetrically between to
copper sites, so a perfectly antiferromagnetic fluctuation would cancel on the
oxygen site [\onlinecite{MMP}].  The result, derived in eq.
(\ref{eq:C-OHyperfine}) is
\begin{equation}
  \frac{1}{^OT_1T} = \frac {C^2} {4}
  \lim_{\omega\to 0} \frac{1}{\omega}
  \sum_q\
  |f(q)|^2 (\chi^{ss''}_q(\omega)+\chi^{aa''}_q(\omega))
  \label{eq:1/OT1T}
\end{equation}
This is precisely the usual result [\onlinecite{MMP}].
Again using the approximation
of eq.~(\ref{eq:k->omega}), we get
\begin{equation}
  \frac{1}{^OT_1T} = C^2 (2T)^2[ \int^\infty_{{\omega_-(T)/2T}}dx
\frac{2x^2-(\omega_+/2T)^2}
  {\sinh^2(x)} +\int^\infty_{{\omega_-(T)/2T}}dx
\frac{6x^2-2(\omega_+/2T)^2-(\omega_+/2T)^2}
  {\sinh^2(x)}]
  \label{eq:oT1}
\end{equation}
In contrast to the yttrium, where the variation was by a factor of two, here
the coefficient of the $T^2$ term varies by a factor of four between the
coupled and decoupled-plane critical regimes.

In the mean field theory the oxygen rate varies as $T^2$ in the critical
regime.  We shall see in the next section that in the more physically
reasonable scaling analysis the antiferromagnetic spin fluctuations lead to an
oxygen relaxation rate proportional to $T$ in the critical regime.

We now turn to the copper rate, which is given by
\begin{equation}
  \frac{1}{{^{Cu}T}_1T}
  =\frac{1}{4}
  \lim_{\omega \to 0}
  \frac {1} {\omega}
  \sum_q (A-4B\gamma(q))^2 (\chi^{ss''}(q,\omega) +\chi^{aa''}(q,\omega))
  \label{eq:CuT1T}
\end{equation}
Again this may be evaluated, giving
\begin{equation}
  \frac{1}{{^{Cu}T}_1T} =
  \frac {(A-4B)^2} {4}
  \left[
\int^\infty_{{\omega_+(T)/2T}}dx/\sinh^2(x)+3\int^\infty_{{\omega_-(T)/2T}}dx/\sinh^2(x)
\right]
\label{eq:cuT1}
\end{equation}
This also is plotted in fig. 4 for the parameters used in fig. 3, and again
there is a factor of four change from the coupled-planes critical regime to
the decoupled-planes critical regime. Note that in the critical regime the Cu
relaxation rate is $T$-independent. Again, this is due to the artificiality of
the Schwinger-boson mean field theory. The scaling analysis predicts a $1/T$
behavior [\onlinecite{Sachdev}]

\section{Scaling Analysis}
\label{sec:scaling}
The mean field analysis of the previous sections is known
[\onlinecite{Sachdev,Chubukov,Chakravarty}] to give incorrect results for
dynamical susceptibilities for the single-plane Heisenberg model at finite
spin degeneracy $N$.  A scaling theory which corrects these discrepancies has
been developed [\onlinecite{CHN,Sachdev}].  In this section we extend the
scaling theory to the coupled-plane system of interest here, and also
construct a universal amplitude ratio for the NMR $T_1$ and $T_2$ relaxation
times.

In the scaling theory of the one-plane model the important parameters for the
physics on the disordered side of the phase boundary are the spin-wave
velocity, $v$, the $T = 0$ gap to spin-1 excitations, $\Delta$, the
quasiparticle residue of the lowest-lying S=1 excitation at $T=0$, $A_{qp}$ and
the
temperature, $T$.  Low energy physical quantities are universal functions of
these parameters. Further, one must distinguish the "quantum disordered" $T
\ll \Delta$ regime from the "quantum critical" $T \gg \Delta$ regime.

We now extend the theory to two coupled planes.  We expect on general grounds,
and showed explicitly using the mean field theory, that the between-planes
coupling $J_2$ splits the spin excitation spectrum into acoustic and optic
branches.  The minimum gap to optic excitations, $\Delta_o$, is nonzero for
$J_2>0$; the acoustic excitations acquire a gap $\Delta_a$ in the disordered
phase.  In the mean field treatment of the previous section,
$\Delta_a=2\omega_+$ while $\Delta_o=\omega_++\omega_-$.  We found that on the
phase boundary $\Delta_o \sim J_2$ while $\Delta_a=0$.  As one moves into the
disordered phase, $\Delta_a$ rapidly approaches $\Delta_o$.  Deep in the
ordered phase, for small $J_2$, $\Delta_o \sim \sqrt{J_2J_1}$ in agreement
with spin-wave theory [\onlinecite{Tranquada}]

As we tune the system through the order-disorder transition, $\Delta_o$
remains nonzero, so the optic fluctuations are irrelevant in the
renormalization group sense.  Within the mean field approximation the
transition is thus one in which the Heisenberg order vanishes and a single
two-fold degenerate spin-wave mode of velocity $v_a$ acquires a gap
$\Delta_a$.  The usual universality arguments then imply that the transition
is in the previously considered universality class, and that at $\omega ,T \ll
\Delta_o$ the physical quantities are given by the previously calculated
universal functions [\onlinecite{Sachdev}] evaluated at arguments $v_a$ and
$T/\Delta_a$.  We refer to this $\omega ,T \ll \Delta_o$ regime as the {\em
coupled-plane} regime, and distinguish the $T \ll \Delta_a$ ``{\em
coupled-plane disordered regime}'' from the $T \gg \Delta_a$ ``{\em
coupled-plane critical regime}''.  Of course, if
$(\Delta_o-\Delta_a)/\Delta_o$ is too small the coupled-planes critical regime
may not exist.

There is one subtlety in the analysis of the coupled-planes regime: the
universal forms give susceptibilities in units of $emu/area$.  A unit cell
contains two Cu atoms (one in each plane); the susceptibilities are evenly
divided between the two planes, thus the susceptibilities per Cu are one half
of the one-plane values.  Further, because the Cu and O nuclei of interest for
high-T$_{\rm c}$ NMR experiments sit in one $CuO_2$ plane, the hyperfine
coupling of the Cu or O nucleus to the surviving spin-degree of freedom is
half as large as in a single-plane model, so the Cu and O relaxation rates,
which go as the square of the hyperfine coupling constant, will be one quarter
of the usual size.  The yttrium nucleus, however, sits between two planes and
over the center of a plaquette of four Cu nuclei.  In the coupled-planes
regime the hyperfine coupling per moment is half the expected value, but the
eight nearest neighbor spins add coherently , so the yttrium NMR rate is only
one half of the expected value.  These factors were explicitly derived in the
mean field analysis of the previous section; we have given a qualitative
argument here.

As one increases the temperature from zero in the disordered phase one passes
first through the coupled planes disordered regime and then, if
$(\Delta_o-\Delta_a)/\Delta_o$ is sufficiently large, through the
coupled-planes critical regime.  As one continues to increase the temperature
it becomes comparable to the optic mode energy $\Delta_o$ and the bilayers
become uncorrelated.  If $\Delta_o$ is sufficiently small, i.e. if $J_2 \ll
J_1$ then at $T \Delta_o$ the physics will still be controlled by a $T=0$
critical point.  We showed using the mean field theory that in this regime
each plane fluctuates independently and is in the quantum critical regime of a
one-plane model.  We believe this conclusion survives beyond mean field
theory.  We refer to the regime $T\Delta_o$ as the ``{\em decoupled-planes
critical regime}''.  It is probable that one could extend the calculation of
reference [\onlinecite{Sachdev}] to incorporate fluctuations into our
mean-field analysis of the crossover between the coupled and decoupled planes
regimes, but we have not attempted this.  Instead, we consider the uniform
susceptibility and NMR in each plane separately. The scaling properties of the
single-plane transition have recently been elegantly derived and discussed
[\onlinecite{Sachdev}].  We summarize, as briefly as possible, the results we
will need and their extension to the two-plane system.

The correlation length , $\xi$ is a universal
function of $T$ and $\Delta$,
\begin{equation}
  \xi=\frac{\hbar v} {k_B T} X(k_BT/\Delta)
  \label{eq:xi}
\end{equation}
with $X(y)$ a universal function which tends a number very nearly  unity (1.03
in a large N expansion with terms of order 1 and 1/N included and N set equal
to 3) as $y$ tends to infinity
(so $\xi \sim \hbar v/k_BT$ for $T \gg \Delta$) and to $y$ as $y$ tends to zero
(so
$\xi$ tends to $\hbar v/\Delta$ for small $T$).  Note that for $J_2 \ll J_1$,
$v$ is
essentially the same for acoustic and optic modes, so that this relation is
valid in both coupled and decoupled-plane regimes.

The divergent part of the order parameter susceptibility, $\chi _{AF}$, may be
written
\begin{equation}
  \chi_{AF}(q,\omega;T) = \chi_{AF} \xi^{2-\eta} \phi_{AF}(q \xi,\omega \xi
  /v;T/\Delta)
  \label{eq:chiaf}
\end{equation}
Here $\chi _{AF}$ contains the dimensions  and
$\phi_{AF}$ is a universal function.  We have introduced the exponent $\eta$
for completeness even though for the transition in question it is very nearly
zero [\onlinecite{Sachdev,eta}].  In the coupled-planes regime we should
interpret this as a susceptibility per unit area; in the decoupled-planes
regime as a susceptibility per area per plane.  Here $\chi_{AF}$ is a critical
amplitude; its value is not universal but from it and other measurable
quantities universal amplitude ratios may be constructed.  In particular,
$\chi_{AF} = A_{qp}/v^2$ where
$A_{qp}$ is the quasiparticle residue of the lowest-lying S=1 excitation in the
disordered phase at T=0 and we have made explicit the factors of the velocity
which were set to unity in [\onlinecite{Sachdev}].

The uniform susceptibility has also a scaling form; here even the dimensional
prefactor is known [\onlinecite{Sachdev}].  The susceptibility in units of emu
per area is
\begin{equation}
  \chi_u(q,\omega;T)=
  \frac {g^2 \mu_B^2} {\hbar v \xi} \phi_u(k_BT/\Delta) \frac{(D_s/\xi)
  (q \xi)^2} {-i \omega \xi + D_s/\xi (q \xi)^2}
  \label{eq:chi-scale}
\end{equation}
Here $g$ is the electron g-factor and $\mu _B$ is the Bohr magneton.  Again
this is a susceptibility per unit area in the coupled planes regime and a
susceptibility per area per plane in the decoupled planes regime.  $D_s$ is
the spin diffusion coefficient; it has a dependence on $T$ and $q$ which we
have suppressed here.  $\phi_u$ is a universal function which tends to a
constant as $y$ tends to infinity and vanishes rapidly as y tends to zero.  It
is usually assumed that the quantity $g \mu_B$  may be taken to have the
free electron values because in the nonlinear sigma model treatment of the
critical point it is not renormalized from its bare value; however this
assumption has not been proven for more general models.

We shall be most interested in applying this formula in the critical regime $T
\gg \Delta$; we therefore proceed to estimate $D_s$ in this regime.  We argue
that for the modes relevant to NMR experiments,
\begin{equation}
  D_s = D_s^0 v^2 \frac {\hbar \ln^{1/2}(1/q)} {k_B T}
\end{equation}
Here $D_s^0$ is a number, presumably of order unity.  The factor of $1/T$
arises as follows.  For excitations of velocity $v$ and scattering rate
$\Gamma$, $D_s \sim v^2/\Gamma$.  Further, conventional dynamic scaling
suggests that modes at scales shorter than the correlation length are weakly
damped, while those at longer scales are overdamped, implying that the
scattering rate for spin excitations is proportional to $T$.  The factor of
$\ln^{1/2}$ comes from the breakdown of hydrodynamics in two spatial dimensions
[\onlinecite{Forster}].

We are now able to discuss relaxation rates.  We begin with the yttrium rate.
We saw in the previous section that the yttrium nucleus is coupled only to
fluctuations of the uniform magnetization.  By comparing the notations of this
and the previous section we see that in the coupled-planes regime the coupling
constant is $4Da^2/\mu_Bg$ (recall that there are 8 Cu neighbors but that the
spin density is evenly divided between the two planes).  Calculating the
relaxation rate in the usual way gives
\begin{equation}
  \frac {\hbar} {^YT_1k_BT} = \frac {16 D^2 a^4} {\hbar^3 D_s^0 v^3 \xi}
\frac{k_BT} {\pi} \ln^{1/2}(1/qa)
\end{equation}
Recall that $\phi_u$ refers to two planes; normalizing per plane restores the
factor of 32 from [\ref{eq:yT1}].  A power of $\ln^1$ was obtained previously
by Chakravarty and Orbach in a calculation of relaxation in an ordered magnet
[\onlinecite{Chakravarty}]; the different power comes because they did not
consider the corrections to hydrodynamics.  The logarithm will be cut off at
some scale by a three dimensional coupling $J_{3D}$ and is presumably not
important in practice.  The important result is that in the quantum disordered
regime both $D_s$ and $\xi$ go as $1/T$ so the yttrium relaxation rate in this
model is proportional to $T^2$ up to logarithms, as was found in the mean
field theory of the previous section.  In the decoupled-planes critical regime
the calculation is identical except that we must add two contributions, one
from each plane.  For each contribution the coupling constant is
$4Da^2/\mu_Bg$ and we must neglect interplanar correlations.  The result is a
factor of two increase in the coefficient of the $T^2$ term, as was found in
mean field theory.

We now consider the oxygen relaxation rate.  This has one contribution from
the small-$q$ fluctuations which may be evaluated as we did for yttrium and
which will be seen to be sub-dominant, and another contribution from the
antiferromagnetic fluctuations, which we evaluate.  The coupling constant
connecting the oxygen nucleus to an antiferromagnetic fluctuation in a given
plane of wavevector $q$ (measured from $P$) may be written $C_{AF} a^2 f(qa)$
where $C_{AF}$ is a priori not the same as the coupling constant $C$
introduced before.  Symmetry implies that $f(x) \sim x^2$ at small $x$.
Combining this with eq. (\ref{eq:chiaf}) for $\chi_{AF}$ gives, in the
coupled planes regime:
\begin{equation}
  \ \frac{\hbar} {^OT_1k_BT} =\frac{1}{4} \frac{C_{AF}^2a^3} {\hbar v}
\frac{\chi_{AF}} {\mu_B^2}
  (\xi/a)^{-(1+ \eta)} \phi _0 (T/\Delta_a)
  \label{eq:orelax}
\end{equation}
Here $\phi _0 (z)$ is a universal function obtained by integrating $\lim_{y
\to 0}x^2 \phi ^{''}_{AF}(x,y;z)/y$ over $x$.   We have assumed the integral
converges; in the spin-only model this is reasonable because at momentum scales
larger than the inverse correlation length the model goes over to spin-wave
theory and the integra
ls there converge.  In a more general model the issue of convergence is less
clear.  Thus the oxygen relaxation rate
in this model scales at $T^{1+ \eta}$ with a non-universal prefactor involving
both the hyperfine coupling $C_{AF}$ and the amplitude $\chi_{AF}$.  The
oxygen relaxation rate scales differently from the yttrium because the oxygen
is coupled (albeit weakly) to the antiferromagnetic fluctuations, while the
yttrium is not.  As we have previously argued, the constant $C_{AF}$ is larger
by a factor of two in the decoupled-planes regime than it is in the
coupled-planes regime, leading to a factor-of-four change in the relaxation
rate.

We finally consider the copper.  This is coupled to the spins by a matrix
element $A_{AF}a^2/ \mu _B$.  The relaxation rate in the coupled-planes regime
is
\begin{equation}
  \frac {\hbar} {^{Cu}T_1k_BT} = \frac {A_{AF}^2 a^3} {4 \hbar v} \frac
  {\chi_{AF}} {\mu _B ^2} (\xi/a) ^{1- \eta} \phi _{Cu} (T/ \Delta)
  \label{eq:curelax}
\end{equation}
Thus the Cu relaxation rate $1/T_1T$ in this model scales at $T^{\eta -1}$
times a non-universal prefactor involving both the hyperfine coupling and the
amplitude $\chi_{AF}$.  The formula [\ref{eq:curelax}] has been previously
given [\onlinecite{Sachdev}].  The same factor of four change in the
coefficient of the leading $T$-dependent term between the coupled-plane and
decoupled-plane regimes that occurred for the oxygen relaxation rate occurs for
the copper.

The $T_1$ relaxation rate is determined by the imaginary part of $\chi$.  The
$T_2$ rate measures the real part of $\chi$.  Specifically, in circumstances
relevant to experiments on high-T$_{\rm c}$ materials
[\onlinecite{Pennington}]
\begin{equation}
  (\frac {1} {T_2})^2 = n_m \sum_q\ [A_{AF} \chi^{'}(q,\omega =0)]^2
\end{equation}
where $n_m$ is the density of NMR nuclei.  Substituting the scaling ansatz and
integrating gives
\begin{equation}
  \frac{\hbar} {T_2} =n_m^{1/2} \frac {A_{AF}^2a^3 \chi_{AF}} {\mu_B^2} (\xi
/a)^{1-\eta}
  \phi _{T2}(T/\Delta)
  \label{eq:T2}
\end{equation}
By combining eqs. (\ref{eq:curelax},\ref{eq:T2}) we see that apart from the
factor $n_m^{1/2}$ and the velocity $v$, the ratio of $T_2$ to $T_1T$ is
universal, and indeed takes the same value in the coupled-planes and
decoupled-planes critical regimes.  Note however that in the coupled-planes
regime the contribution of the optic excitations to $1/T_2$ will be large (of
order $1/\Delta_o$).  The contribution of the acoustic sector is of order
$1/\Delta_a$.  Thus $1/T_2$ will attain its universal value in the
coupled-planes regime only if $(T,\Delta_a) \ll \Delta_o$

Sokol and Pines [\onlinecite{Sokol}] have previously made the interesting
observation that the observed $T$-independence of $T_2/T_1T$ in
YBa$_2$Cu$_3$O$_{6.6}$ for $T>150\ {\rm K}$ suggests that the magnetic
dynamics in this material is controlled by the $z=1$ critical point considered
here.  We
see that the magnitude provides information about the velocity, $v$, and that
this must be consistent with the uniform susceptibility.

\section{Conclusion}

We have studied some aspects of the $T=0$ magnetic-non-magnetic transition
occurring in a model of two antiferromagnetically coupled planes of
antiferromagnetically correlated spins. The model is defined in eq.
(\ref{eq:Model}) and depicted in fig. 1.  The two dimensionless parameters are
$J_2/J_1$ (the ratio of the between-planes coupling $J_2$ to the in-plane
coupling $J_1$) and $S$, the magnitude of the spin in one plane.  The
important dimensional parameter is the spin wave velocity, $v$.  The phase
diagram at $T=0$ in the $J_2/J_1$ - $S$ plane is shown in fig. 2.  For a
single plane (i.e. $J_2=0$), decreasing $S$ through a critical value $S_c$
causes a phase transition between a magnetically ordered phase and a singlet
phase with a gap to spin excitations.  Some properties of this transition were
determined by Chakravarty, Halperin and Nelson [\onlinecite{CHN}] and it was
analyzed in detail by Sachdev, Chubukov and Ye[\onlinecite{Sachdev}]. For the
coupled-plane system we found using a Schwinger-boson mean field theory that a
large value of the interplanar coupling $J_2$ destroys the magnetism even for
$S S_c$ (because it favors binding of nearest neighbor spins on different
planes into singlets), while a small $J_2$ promotes order by increasing the
effective size of the spin in a unit cell.  The interplay of these two
different sorts of physics leads to the reentrant phase diagram shown in fig.
2.  This phase diagram differs in an important respect from our previous
interpretation of the data on spin susceptibilites of La$_{\rm
2-x}$Sr$_x$CuO$_4$ and YBa$_2$Cu$_3$O$_{6.6}$ [\onlinecite{MM PRL}]. In both
compounds it is clear that doping destroys the magnetism.  In La$_{\rm
2-x}$Sr$_x$CuO$_4$ interplane coupling is negligible and at least the Cu
relaxation rate and single crystal susceptibility data show no clear evidence
of a singlet phase with a gap to excitations (although some susceptibility and
Knight shift
data have been so interpreted [\onlinecite{Sokol,Sachdev}]).  On the other
hand, in
YBa$_2$Cu$_3$O$_{7-\delta}$ it is clear that the nearest-neighbor CuO$_2$
planes are coupled by an interaction at least of order 300 K, and "spin-gap"
effects are very easily observed in susceptibilities and relaxation rates for
$0.1 < \delta < 0.5$.  Thus it appears that in the real materials a presumably
modest between-planes coupling promotes "spin-gap" behavior, and therefore
that the spin-only model is missing some essential feature of the physics,
most likely related to the presence of mobile holes.

Although it is not completely realistic, the coupled-plane model might capture
some aspects of the physics of YBa$_2$Cu$_3$O$_{6.6}$ and YBa$_2$Cu$_4$O$_8$.
We therefore calculated the predictions of the model for the temperature and
frequency dependence of the susceptibilities measured in NMR and neutron
scattering using the Schwinger boson mean field method and a scaling analysis.
We studied parameters such that the model has no long range order at $T=0$.
The behavior in the disordered phase of a single plane of Heisenberg spins is
understood [\onlinecite{Sachdev,Sachdev and Read}].  In the single-plane case,
the important parameter is the $T=0$ gap to spin one excitations, $\Delta$.
The spin-one excitations are essentially spin waves with a gap.  There are
several regimes of temperature.  For $T \ll \Delta$ the number of thermal spin
excitations is negligible; the static spin
susceptibility at $q=0$ and the dissipative part of the dynamic susceptibility
at all $q$ and at $\omega \ll \Delta$ have an activated temperature dependence
$ \sim e^{- \Delta/T}$.  This regime is referred to as the "quantum disordered
regime".  If the microscopic exchange constant $J \gg \Delta$
then for $\Delta \ll T \ll J$ another regime exists in which the physics is
dominated by the $T=0$ critical point but the gap is not important.  In this
regime the static uniform susceptibility is proportional to $T$ and the
antiferromagnetic correlation length grows as $1/T$.  The regime is referred to
as the "quantum critical regime".

In the two plane model of interest here the between-planes coupling $J_2$
splits the spin excitation spectrum into acoustic and optic modes.  There are
two important scales: $\Delta_a$, the $T=0$ gap to acoustic excitations and
$\Delta_o$, the $T=0$ gap to optic excitations.  Both gaps are nonzero in the
disordered phase.  At the antiferromagnetic-singlet transition $\Delta_a$
vanishes
while $\Delta_o$ remains non-zero.  At the transition we found from the mean
field theory that $\Delta_o =J_2$.  As one moves deeper into the disordered
phase, $(\Delta_o-\Delta_a)/\Delta_o$ decreases rapidly.  If one is
sufficiently close to the phase boundary, so that $0 <
\Delta_a \ll \Delta_o \ll J_1$, there are three regimes.  These are depicted in
fig. 6.  For $(T,\omega ) \ll \Delta_o$ the optic mode of the two plane system
is frozen out and the physics is dominated by the acoustic mode of the two
plane system.  We refer to this as the "coupled-planes" regime.  For
$(T,\omega ) \ll \Delta_a$ even the acoustic mode is frozen out.  This is the
"coupled-planes disordered regime".  For $\Delta_a \ll T \ll \Delta_o$ the
acoustic mode is thermally activated and the system is in the "coupled-planes
quantum critical regime".  Finally, for $\Delta_o \ll T$ the coupling between
the planes become negligible and the planes fluctuate independently.  This is
the "decoupled planes" regime.  If $\Delta_o \ll T \ll J_1$ then we argued
that the spin dynamics in each plane is separately given by the quantum
critical behavior of a one-plane model.  Of course if $(\Delta_o
-\Delta_a)/\Delta_o$ is too small, the coupled-plane critical regime does not
exist, while if $\Delta_o/J_1$ is too large, the physics in the
decoupled-planes regime will not be controlled by a $T=0$ critical point.

We now summarize the results we obtained in sections III and IV for the
temperature and frequency dependence of the NMR rates and susceptibilities.
We found that the difference, $\Delta_o-\Delta_a <J_2$ everywhere in the
disordered phase, and that the contribution from the optic modes near the
antiferromagnetic point was rather small and only weakly temperature dependent
if $\Delta_o \gg \Delta_a$.  The uniform susceptibility, $\chi(T)$, is
activated in the coupled planes disordered regime; in the coupled-planes
critical regime $\chi(T) =0.5ET$ (here $E$ is a number) and in the
decoupled-planes critical regime $\chi(T) = ET$.  The factor of two change in
the coefficient of $T$ between the coupled and decoupled planes regimes was
derived explicitly from the Schwinger boson mean field theory.  The physical
origin, we believe, is that in the coupled-plane regime one of the two spin
modes at each k (namely, the optic mode) is frozen out and does not contribute
to $\chi$, whereas in the decoupled-planes regime both modes contribute.

The $T$-dependences of the nuclear relaxation rates are more subtle.  They are
summarized in fig 6. We consider
first the copper and oxygen rates.  In the quantum disordered regime all rates
are activated.  In the coupled-planes critical regime the Cu rate $1/T_1T
=0.25 A/T$ while the oxygen rate is given by $0.25CT$ Here we have set the
exponent $\eta$, which is in practice very small [\onlinecite{eta}] to
zero.  $A$ and $C$ are constants.  In the decoupled planes critical regime the
formulae are the same except that the number 0.25 becomes 1.  The factor of
four change in the coefficient of the leading $T$-dependence of the Cu and O
rates between the coupled-planes and decoupled-planes critical regimes was
derived from the mean field theory.  We believe that the physical origin is
that in the coupled-planes regime only the acoustic mode contributes to
relaxation rates.  In this mode the spin fluctuation is even divided between
the two planes; the hyperfine coupling constant connecting a Cu or O nucleus
in a given plane to the spin fluctuation is thus half of what it would be in a
one-plane theory, and the rate goes as the square of the hyperfine coupling.
The factor of four variation has an interesting implication for estimates of
the relative strengths of the antiferromagnetic spin fluctuations in different
materials.  Authors (including us) who had considered the question previously
argued that because the Cu relaxation rate in YBa$_2$Cu$_3$O$_{7-\delta}$ was
smaller than in La$_{\rm 2-x}$Sr$_{\rm x}$CuO$_4$, while the hyperfine
couplings were approximately the same, the spin fluctuations must be weaker in
the former material.  We see now that until one knows whether YBa$_2$ Cu$_3$
O$_{7-\delta}$ is in the coupled-planes or decoupled-planes
regime one cannot meaningfully compare the magnitudes of the relaxation rates
to those of $La_2CuO_4$.

The yttrium relaxation relaxation rate behaves slightly differently, because
the yttrium nucleus is coupled to a symmetric combination of Cu nuclei in the
two planes.  In the disordered regime the yttrium rate is activated, in the
coupled-planes critical regime it goes as $0.5DT^2$ and in the
decoupled-planes critical regime it goes as $DT^2$.  The yttrium rate changes
only by a factor of two between the coupled-planes and decoupled-planes
critical regimes because in the coupled-planes regime the spins in different
planes move coherently while in the decoupled-planes regime the spins move
incoherently. The difference between coherent and incoherent addition of spin
fluctuations produces a factor of two which partially compensates for the
factor of four discussed previously.

We now consider the implications of our results for experiments on
YBa$_2$Cu$_3$O$_{6.6}$ and YBa$_2$Cu$_4$O$_8$. Below $T=T^* \sim 150\ {\rm K}$
all of the relaxation rates including the Cu $1/T_1T$ drop as $T$ is decreased.
We believe that this can only occur if the physics below $T \sim 150\ {\rm K}$
is dominated by thermal excitations above a $T=0$ singlet state
with a gap to spin excitations. In a spin-only model such as that considered
here one would model this by choosing a value of spin $S$ such that the system
was near to, but on the disordered side of, the phase boundary in fig. 2.
Further, the value of $T^*$ implies that $\Delta_a \sim 150\ {\rm K}$. We must
next consider the value of the interplanar coupling $J_2$.  The experimental
evidence is not conclusive.  Only the acoustic excitation of the two-plane
system has been observed via neutron scattering in any member of the YBa$_2$
family of high-T$_{\rm c}$ materials [\onlinecite{Tranquada,Mook}]. The main
effort has been at energies less than 40 meV and temperatures less than 150\
{\rm K}.  This would suggest that in the metallic YBa materials, $\Delta_o
\ge 40\ {\rm meV}$.  NMR measurements on YBa$_2$Cu$_4$O$_8$ provide more
information.  The Cu $T_1$ has been found to obey very well the Curie law
$1/T_1T \sim 1/T$ for $200\ {\rm K} < T < 700\ {\rm K}$ [\onlinecite{Machi}].
This implies that the temperature scale at which the bilayers become decoupled
in this material
is less than 200\ {\rm K} or greater than 700\ {\rm K}.  We believe a bilayer
coupling greater than 700 K would be very hard to justify on theoretical
grounds.  The argument is that the in-plane $J \sim 1500\ {\rm K}$, while $J_2
\ll J_1$ because band structure [\onlinecite{Pickett}] and photoemission
[\onlinecite{Argonne}] results imply that the between-planes hybridization is
much less than the in-plane hybridization, and the exchange energy scales as a
high power of the hybridization. However, a bilayer coupling much less than
200 K may not be consistent with the neutron data.  From the mean field theory
we found $\Delta_o \le J_2+\Delta_a$; our estimate $\Delta_a \sim 150\ {\rm
K}$ then implies $\Delta_o \le 350\ {\rm K}$ if $J_2 < 200\ {\rm K}$.  Thus
presently available data provide somewhat contradictory answers to the
question whether YBa$_2$Cu$_4$O$_8$ is in the coupled-planes or
decoupled-planes regime for $200\ {\rm K} \le T \le 700\ {\rm K}$.  In what
follows we consider both possibilities.

We now turn to a more quantitative discussion.  Sokol and Pines have proposed
that the magnetic dynamics of underdoped high-$T_c$ materials are determined by
the critical point discussed here, and their discussion has been amplified and
extended by Barzykin, Pines, Sokol and Thelen [\onlinecite{Sokol}].  They do
not consider the bilayer coupling.  They propose that these materials are in
the quantum critical regime for $T \ge 150\ {\rm K}$ and in the quantum
disordered regime for $T \le 150\ {\rm K}$.  The essential piece of evidence
they cite in favor of their proposal is the observed approximate
$T$-independence of the ratio $T_1T/T_2$ for $150\ {\rm K} \le T \le 300\ {\rm
K}$.  The observed [\onlinecite{Takigawa3}] magnitude of this ratio is
approximately one-half of the value observed [\onlinecite{Imai}] in
La$_2$CuO$_4$ and calculated [\onlinecite{Singh}] for the $S=1/2$ Heisenberg
model with $J=0.13\ {\rm eV}$.  The density $n_m$ of $^{63}Cu$ NMR ions is the
same in
the La-Sr and YBa materials, so we conclude from eq. (\ref{eq:T2}) that if the
magnetic dynamics of YBa$_2$Cu$_3$O$_{6.6}$ and YBa$_2$Cu$_4$O$_8$ are well
described by the universal scaling forms in either the coupled-planes or the
decoupled-planes regimes, then the appropriate spin-wave velocity $v$ is about
one half of the $0.8\ {\rm eV}\;\AA$ appropriate for La$_2$CuO$_4$
[\onlinecite{Aeppli}], .i.e.
\begin{equation}
v_{\rm YBa} = 0.4\ [{\rm eV}\;\AA]
\label{eq:vyba}
\end{equation}
The value of $v$ makes a prediction for the magnitude of the correlation
length.  From eq. [\ref{eq:xi}] we find $\xi/a = 4$ at $T=300K$ and $\xi/a =6$
at $T=200K$; below this temperature the crossover to the quantum disordered
regime presumably means that the T dependence of the correlation length becomes
much weaker.  These lengths are rather larger than observed in neutron
scattering experiments [\onlinecite{Rossad,Tranquada}].

These considerations also have an implication for the Cu $T_1$ relaxation
rate.  In the quantum critical regime, the theoretical result for the Cu
relaxation rate is $1/T_1T = A/T$.  The coefficient $A \sim A_{qp}/v^2$
[\onlinecite{Sachdev}] where $A_{qp}$ is the quasiparticle residue of the
lowest lying S=1 state above the gap at $T=0$.  It may in principle be
determined from neutron scattering mea
surements at low T.  The coefficient A also
depends on hyperfine couplings (which have been
claimed to be the same in La$_2$CuO$_4$ as in YBa$_2$Cu$_3$O$_{6.6}$ and
YBa$_2$Cu$_4$O$_8$ [\onlinecite{MMP,Imai2}]) and on whether the material is in
the
coupled-planes or decoupled-planes regimes.  In the absence of a measurement of
$A_{qp}$ one cannot definitively calculate the Cu relaxation rate; however, it
is interesting to attempt to estimate it.  We first argue as follows:
$La_2CuO_4$ has been clai
med to be in the quantum critical regime for $T>600K$
[\onlinecite{Sokol,Sachdev,Imai}]; further, at these temperatures the Cu $T_1$
has only a weak doping dependence, implying that the combination $A_{qp}/v^2$
depends only weakly on doping.  If this is c
orrect, then we would expect that if YBa$_2$Cu$_3$O$_{6.6}$ and
YBa$_2$Cu$_4$O$_8$ were in the decoupled planes regime, A would be well
approximated by the value $A = 3300 \ {\rm sec}^{-1}$ [\onlinecite{Imai2}]
appropriate to La$_2$CuO$_4$, while if it were in the coupled plane regime we
would expect $A = 800 \ {\rm s
ec^{-1}}$.  (In fact, the values might be slightly larger since the $J$ for
$YBa_2Cu_3O_{6.0}$ is smaller than the J for $La_2CuO_4$ [\onlinecite{J}]).  In
fact, in YBa$_2$Cu$_4$O$_8$,  $A = 1600 \ {\rm sec^{-1}}$ [\onlinecite{Machi}]
compatible with neit
her estimate.  An alternative argument would be to say that $A_{qp}$ has the
dimension of energy, and the characteristic energy scale is given by dividing
the velocity, $v$, by the lattice constant, a, implying $A \sim 1/v$ so from eq
[\ref{eq:vyba}] we w
ould expect that in the coupled-planes regime the value of the Cu $T_1$ would
come out about right.  A third possibility is that for some unknown reason
$A_{qp}$ could drop by a factor of 8 upon going from $La_{2-x}Sr_xCuO_4$ to
YBa$_2$Cu$_4$O$_8$, so tha
t t
he data for $T >150K$ would be consistent with the decoupled planes regime.  Of
course a fourth possiblity is that the theory is not applicable.  Neutron
measurements of absolute scattering intensities on underdoped YBa materials
would be very helpful in
resolving this issue.

Our discussion so far has emphasized properties related to the
antiferromagnetic fluctuations.  These are in a sense robust and relatively
model-independent, depending as they do primarily on the existence of a
growing correlation length, a weak $J_2$ and a spin-gap at low $T$.  We now
consider small-$q$ properties.  In the Heisenberg model, the behavior of the
small-$q$ susceptibility is very closely tied to the behavior of the large-$q$
susceptibility [\onlinecite{Sachdev,Forsterbook}].  In a more
fermi-liquid-like model this need not be true, and therefore the bilayer
coupling, which strongly affects the behavior near the antiferromagnetic
point, need not in a more realistic model also strongly affect the
susceptibility near $q=0$.  It is also clear that the spin-only model is not,
by
itself, a reasonable description of the small-$q$ spin dynamics of
$YBa_2Cu_3O_{7-\delta}$ or $YBa_2Cu_4O_8$. For example, the observed yttrium
nuclear relaxation rate does not vary
as $T^2$ but is more nearly proportional to the static susceptibility
[\onlinecite{Takigawa}].  Also, our calculated oxygen nuclear relaxation rate
is too small by about a factor of 16 to explain the data
[\onlinecite{critical}].  Thus, we must invoke an extra contribution to
$\chi^{''}$ existing at least at small q.  It is
 natural to suppose that this is due to the mobile carriers and that thereofre
the contribution to $\chi^{''}$ is of more or less the fermi-liquid form.
However, the fermi-liquid like contribution cannot be appreciable near the
antiferromagnetic point or
 it would overdamp the spin waves and change the universality class of the
magnetic fluctuations [\onlinecite{Hertz,AJM}].  It has very recently been
argued [\onlinecite{comment}] that in the Shraiman-Siggia model of doped
antiferromagnets [\onlinecite{SS
}] precisely the required behavior occurs, with the fermions making an additive
contribution to the small q but not the large q susceptiblities. A similar
conclusion was drawn from high temperature series expansions
[\onlinecite{Singh}].  In the resulting
 tw
o-component picture $\chi=\chi_{spin}+\chi_{qp}$, with the
fermi-liquid like piece $\chi_{qp}$ providing the yttrium and oxygen relaxation
 and the $\chi_{spin}$ given by the theory we have
discussed and providing the nontrivial temperature dependence of the uniform
susceptibility and the Cu relaxation rate. Sokol and co-workers have made a
similar argument on phenomenological grounds, claiming that the
temperature dependence of $\chi$ in YBa$_2$Cu$_4$O$_8$ for $200\ {\rm K} \le T
\le 700\ {\rm K}$ is consistent with the behavior of the spin-only model in
the quantum critical regime [\onlinecite{Sokol}].  Now whether one supposes
the material to be in the coupled-planes or decoupled-planes critical regime,
the theory implies $\chi(T) = a+bT$.  We ignore the value of $a$, on the
grounds that it is dominated by the fermions which are beyond the scope of the
theory, and consider the value of $b$, which is given by eq.
(\ref{eq:chi-scale}).  From eq. (\ref{eq:v
yba}) assuming $g \mu_B$ takes the same value as in $La_2CuO_4$ we find $b=5
\times 10^{-3}\ {\rm states/eV-Cu-K}$ for the
decoupled-planes regime and $b=2.5 \times 10^{-3}\ {\rm states/eV-Cu-K}$ for
the coupled-planes regime.  The data say that  $b \sim 1.6 \times 10^{-3}\ {\rm
states/eV-Cu-K}$ [\onlinecite{Brinkman}].  This too weak $T$
dependence of $\chi$ in the critical regime suggests to us that the
straightforward two-component approach is not applicable to YBa$_2$Cu$_4$O$_8$,
and that the presence of carriers modifies the magnetic behavior more
dramatically. We note, however, that
two results which seem to be
qualitatively consistent with the calculations presented here are: (a)
although the difference is not dramatic, the yttrium rate seems to drop faster
than the oxygen rate as $T$ is lowered and (b) the magnitude of the yttrium
rate is larger than expected from a model in which the planes are uncoupled
[\onlinecite{Takigawa2}].

Whatever is the correct theory, the distinction we have drawn between the
coupled-planes and the decoupled-planes regimes will still be important.
Further, if the NMR and neutron data on underdoped YBa superconductors are both
taken at face value, then t
he crossover between the coupled-planes and decoupled-planes regimes occurs
either at $T>700K$ or at $T \sim 200K$.  The larger value seems to us to
require an implausibly large between-planes coupling; the smaller value would
imply that the crossover to
the quantum disordered regime is complicated by a simultaneous freezing out of
the optic mode of the bilayer system.
\label{sec:conclusions}

\acknowledgements

H. M. was supported in part by the NSF under Grant No. PHY89-04035.  We thank
D. Pines and A. V. Sokol for helpful discussions and for preprints of their
work. A. J. M.
thanks P. C. Hohenberg for a helpful discussion concerning amplitude ratios and
R. E. Walstedt for a critical reading of the manuscript.
The authors thank the Correlated Electron Theory Program at Los Alamos
National Laboratory for hospitality while part of the manuscript was written.

\appendix

\section{Derivation of Mean Field Equations}

We begin from eq.~(\ref{eq:Model}).  We write the model as a functional
integral, introduce a field $Q_{<i,j>}^{(a)}$ to decouple the $J_1$
interaction, a field $\Delta_i$ to decouple the $J_2$ interaction and a field
$\mu$ to enforce the constraint.  We find for the partition function, $Z$,
\begin{equation}
  Z = \int{\cal D}\Delta^+\Delta{\cal D} Q^+Q{\cal D} b^\dagger b{\cal D} \mu
  \exp\left(-\int_0^\beta d\tau{\cal L}'\right)
\end{equation}
with
\begin{eqnarray}
  {\cal L}' &=& \sum_{ia\alpha}
  b^{\dagger(a)}_{i\alpha}[\partial_\tau+\mu_i^{(a)}]b^{(a)}_{i\alpha}\cr &+&
  \frac{1}{4}\sum_{<i,j>,a\alpha}
  b^{\dagger(a)}_{i\alpha}b^{\dagger(a)}_{j\alpha} Q^{(a)}_{<i,j>}\ + h.c.\cr
  &+& \sum_{i\alpha} b^{\dagger(1)}_{i\alpha}b^{\dagger (2)}_{i\alpha}\Delta_i\
  + h.c.\cr &+& \sum_{<i,j>a}\frac{|Q_{<i,j>}^{(c)}|^2}{8J_1} +\sum_i
  \frac{2|\Delta_i|^2}{J_2}
\label{eq:L'}
\end{eqnarray}
The different normalizations of $Q$ and $\Delta$ have been introduced, so that
the final expression for the boson energy, eq. (\ref{eq:A-BosonEnergy}), has
no numerical factors.  We next introduce symmetric (s) and antisymmetric (a)
bose fields via
\begin{mathletters}
\label{eq:Symmetric}
\begin{eqnarray}
  b^{(1)}_{k\alpha} &=&{1\over\sqrt{2}}(s'_{k\alpha}+a'_{k\alpha}) \\
  b^{(2)}_{k\alpha} &=&{1\over\sqrt{2}}(s'_{k\alpha}-a'_{k\alpha})
\end{eqnarray}
\end{mathletters}
We make the mean field approximation of space and time independent $Q, \Delta,
\mu$, Fourier transform the boson operators and obtain
\begin{eqnarray}
  {\cal L_B} &=& \sum_{k\alpha}s^{\prime\dagger}_{k\alpha}[\partial_\tau+\mu]
  s'_{k\alpha}+[{1\over2}
  (Q\gamma_k+\Delta)s^{\prime\dagger}_{k\alpha}s^{\prime\dagger}_{-k\alpha} +
  h.c.]\cr &+&
  \sum_{k\alpha}a^{\prime\dagger}_{k\alpha}[\partial_\tau+\mu]a'_{k\alpha}+
  [{1\over2}(Q\gamma_k-\Delta)
  a^{\prime\dagger}_{k\alpha}a^{\prime\dagger}_{-k\alpha} + h.c.]
\label{eq:LB}
\end{eqnarray}
with
\begin{equation}
  \gamma_k={1\over2}(\cos k_x+\cos k_y)
\end{equation}
The boson part of this equation may be decoupled by a Bogoliubov
transformation.  The resulting quasiparticles $s$ and $a$ are defined by
\begin{eqnarray}
  s^{\prime\dagger}_{k\alpha} &=& \cosh\theta_k s^\dagger_{k\alpha} -
  \sinh\theta_k s_{-k,\alpha} \cr a^{\prime\dagger}_{k\alpha} &=&
  \cosh\theta_{k+P} a^\dagger_{k\alpha} + \sinh\theta_{k+P} a_{-k,\alpha}
\end{eqnarray}
with
\begin{equation}
  \tanh 2\theta_k=[Q\gamma_k+\Delta]/\mu
\end{equation}
and
\begin{equation}
  P=(\pi,\pi)
\end{equation}
The energy of the $s$-bosons is
\begin{equation}
  \omega_k=\sqrt{\mu^2-(Q\gamma_k+\Delta)^2}
\label{eq:A-BosonEnergy}
\end{equation}
The energy of an $a$-boson at a wavevector $k$ is $\omega_{k+P}$.

The free energy $F$ may be computed in the standard way and is
\begin{equation}
  F=4NT\sum_k \ln [2\sinh(\omega_k/2T)]+NQ^2/2J_1+2N\Delta^2/J_2-2N(1+2S)\mu
\end{equation}
The mean field equations, eqs. (\ref{eq:MF}), follow from differentiating this
equation with respect to $\mu$, $Q$ and $\Delta$.

\section{Approximate Solution of Mean Field Equations}

We begin with eqs. (\ref{eq:MF}). We recast them as $\int
\frac{d^2k}{(2\pi)^2} \rightarrow \int d\gamma N(\gamma)$, we replace

$N(\gamma)$ by $1/2$, we normalize $Q$ and $\Delta$ by $\mu$ and integrate,
obtaining at $T=0$

\begin{mathletters}
\label{eq:MF T=0}
\begin{eqnarray}
  \frac{\sin^{-1} (\Delta+Q)-\sin^{-1}(\Delta-Q)} {2Q} &=& 1+2S
\label{eq:MF T=0 a}\\
\frac{\sin^{-1} (\Delta+Q) -\sin^{-1}(\Delta-Q)
+ (\Delta-Q)\sqrt{1-(\Delta+Q)^2}
- (\Delta+Q)\sqrt{1-(\Delta-Q)^2}} {4Q^2}
&=&\frac{Q\mu}{2J_1}
\label{eq:MF T=0 b}\\
\frac{\sqrt{1-(\Delta-Q)^2}  -  \sqrt{1-(\Delta+Q)^2}} {2Q} &=& \frac{2\Delta
\mu}{J_2}
\label{eq:MF T=0 c}
\end{eqnarray}
\end{mathletters}
These three equations may be reduced to one by taking the sine of
eq.~(\ref{eq:MF T=0 a}) and substituting into eq. (\ref{eq:MF T=0 b}) to
obtain an equation for $\mu(Q)$, solving eq.~(\ref{eq:MF T=0 c}) to obtain an
equation for $\Delta(Q)$ and then substituting the results into
eq.~(\ref{eq:MF T=0 b}). However, for our purposes a simpler approach
suffices. We first locate the critical point at which the minimum boson energy
vanishes. In the notation of this appendix this implies $\Delta^*+Q^*=1$ (we
denote by $*$ the values of the quantities at the critical point). Then
eq.~(\ref{eq:MF T=0 a}) may be solved for $Q^*$.  For $S>S_c=(\pi/2-1)/2$ this
has only one solution; e.g. at $S=1/2$
\begin{equation}
Q^* \cong 0.277
\end{equation}
For $S^*<S<S_c$, with $S^*\cong 0.19$ there are two solutions, one at $Q$ near
1 which is the lower energy solution for $J_1 \gg J_2$ and one at $Q$ near 1/2
which is the lower energy solution for $J_1$ near $J_2/4$.  For $S<S^*$ there
are no solutions.

Once a solution for $Q^*$ is found, eq. (\ref{eq:MF T=0 c}) implies
\begin{equation}
  \frac{\mu^*}{J_2} = \frac{1}{2\sqrt{Q^*(1-Q^*)}}
\end{equation}
and eq. (\ref{eq:MF T=0 b}) implies
\begin{equation}
  \frac{J_2^*}{J_1^*} = 2 \frac{(1+2S)\sqrt{Q^*(1-Q^*)}-1+Q^*}{Q^{*2}}
\label{eq:phaseline}
\end{equation}
Solving eq. (\ref{eq:MF T=0 a}) and then using the solution in eq.
(\ref{eq:phaseline}) yields the phase diagram given in fig. 2.

We now consider the $T 0$ behavior.  We are most interested in the regime
near the phase boundary, and in small $J_2$.  We therefore solve the equations
perturbatively in the small parameters $S_c-S$, $J_2/J_1$ and $T$.  We neglect
terms of third order and higher in these small parameters.  To this order the
$T$ dependent terms may be evaluated exactly.  The equations are conveniently
expressed in terms of the variables $\omega_+$, $\omega_-$ and $\mu$ and are
(note we need $\mu$ only to first order in the small parameters)

\begin{mathletters}
\label{eq:MF PT}
\begin{eqnarray}
  \frac {\pi - \omega_+/\mu-\omega_-/\mu
  +(2T/\mu)(f(\omega_+/T)+f(\omega_-/T))} {2(1-\omega_+^2/(4\mu^2)
  -\omega_-^2/(4\mu^2))} &=& 1+2S
  \label{eq:MF PT a}\\
  \frac{\pi -\omega_-/\mu-\omega_+/\mu+(2T/\mu)(f(\omega_+/T)+f(\omega_-/T))}
  {4} &=&\frac{\mu}{2J_1}
  \label{eq:MF PT b}\\
  \omega_--\omega_++2T(f(\omega_+/T)-f(\omega_-/T)) &=&
  \frac{\omega_-^2-\omega_+^2}{J_2}
  \label{eq:MF PT c}
\end{eqnarray}
\end{mathletters}
Here the function $f$ is defined by
\begin{equation}
  f(x) = -\ln[1-e^{-x}]
\label{eq:df f}
\end{equation}
We use eq. (\ref{eq:MF PT b}) to solve for $\mu$.  Substituting and
rearranging gives
\begin{mathletters}
\begin{eqnarray}
  \omega_++\omega_--2T(f(\omega_+/T)+f(\omega_-/T)) -
  \frac{\omega_+^2+\omega_-^2} {j} &=& \epsilon \\
  \omega_--\omega_+-2T(f(\omega_-/T)-f(\omega_+/T)) &=&
  \frac{\omega_-^2-\omega_+^2}{J_2}
\end{eqnarray}
\label{eq:SmallJ2}
\end{mathletters}
with
\begin{equation}
  j=2J_1(1+8(S_c-S)/\pi)
\end{equation}
and
\begin{equation}
  \epsilon = \frac{2\pi J_1 (S_c-S)}{1+8(S_c-S)/\pi}
\end{equation}
These two equations may be easily solved numerically for $\omega_+$ and
$\omega_-$ by adding the two equations to obtain an expression for $\omega_+$
in terms of $\omega_-$ and $T$, and then substituting that into one of the two
equations to get a single equation for $\omega_-$.  At low $T$ and
sufficiently close to the phase boundary the equations have two solutions, one
with $\omega_+ < \omega_-$ and one with $\omega_+ = \omega_-$; above a
critical temperature the two solutions merge.  By substituting the results in
to eq. (\ref{eq:A-BosonEnergy}) we have verified that where the solution with
$\omega_+ < \omega_-$ exists it has a lower energy than the solution with
$\omega_+ = \omega_-$.  The results displayed in fig. 3 were obtained in this
manner.

\section{Susceptibilities and Relaxation Rates}

We begin with the dynamic susceptiblities, which we obtain by computing the
linear response of the system to an externally applied magnetic field $\vec
h_i^{(a)}$.  The Schwinger boson formalism is rotationally invariant. We
therefore compute only the response to a field in the $z$-direction. Thus we
add to the Hamiltonian a term

\begin{equation}
  \Delta H=\sum_{ia}h^{z(cc)}_i\cdot S_i^{(a)}
\end{equation}
After using eq.~(\ref{eq:Spin}), eq.~(\ref{eq:Staggering}), and
eq.~(\ref{eq:Symmetric}) this becomes
\begin{equation}
  \Delta H =\sum_q
  {h^{(1)}_q-h^{(2)}_q\over2}O^a_q+{h^{(1)}_q+h^{(2)}_q\over2}O^S_q
\end{equation}
with
\begin{mathletters}
\begin{eqnarray}
  O^a_q &=& \sum_{k\alpha\beta}\cosh(\theta_{k+q+P}+\theta_k) (
  s^\dagger_{k+q+P\alpha} \sigma^z_{\alpha\beta} s_{k\beta} +
  a^\dagger_{k+q+P\alpha} \sigma^z_{\alpha\beta} a_{k+q\beta}) \cr &+&
  \sinh(\theta_{k+q+P}+\theta_k) (
s^\dagger_{k+q+P\alpha}\sigma^z_{\alpha\beta}
  s^\dagger_{-k\beta} + h.c. + a \to s) \\ O^s_q &=&
  \sum_{k\alpha\beta}\cosh(\theta_{k+q}-\theta_k) ( s^\dagger_{k+q\alpha}
  \sigma^z_{\alpha\beta} a_{k+P\beta} + a^\dagger_{k+q\alpha}
  \sigma^z_{\alpha\beta} s_{k+p\beta} + h.c.) \cr &+&
  \sinh(\theta_{k+q}-\theta_k) ( s^\dagger_{k+q\alpha} \sigma^z_{\alpha\beta}
  a^\dagger_{-k-P\beta} + a^\dagger_{k+q\alpha} \sigma^z_{\alpha\beta}
  s^\dagger_{-k-\beta} + h.c.)
\end{eqnarray}
\end{mathletters}
The only non-zero correlation functions are
\begin{mathletters}
  \begin{eqnarray}
    \chi^{aa}_q(\omega) &=&\int^\infty_0 dt e^{i(\omega+i\epsilon)t}
    <{[O^a_q(t),O^a_{-q}(0)]}>\cr &=& 4 \sum_k \cosh^2(\theta_{k+q+P}+\theta_k)
    \frac{ b(\omega_k)-b(\omega_{k+q+P}) }{
    \omega-\omega_k+\omega_{k+q+P}+i\epsilon }\cr &+& 4 \sum_k
    \sinh^2(\theta_{k+q+P}+\theta_k) \frac{ [1+b(\omega_k)+b(\omega_{k+q+P})]
    (\omega_k+\omega_{k+q+P}) }{
    (\omega_k+\omega_{k+q+P})^2-(\omega+i\epsilon)^2 }
\end{eqnarray}
and
\begin{eqnarray}
  \chi^{ss}_q(\omega) &=&\int^\infty_0 dt e^{i(\omega+i\epsilon)t}
  <{[O^S_q(t),O^S_{-q}(0)]}>\cr &=&4 \sum_k \cosh^2(\theta_{k+q}-\theta_k)
  \frac{ b(\omega_k)-b(\omega_{k+q}) }{ \omega-\omega_k+\omega_{k+q}-i\epsilon
  }\cr &+& 4 \sum_k \sinh^2(\theta_{k+q}-\theta_k) \frac{
  [1+b(\omega_k)+b(\omega_{k+q})] (\omega_k+\omega_{k+q}) }{
  (\omega_k+\omega_{k+q})^2-(\omega+i\epsilon)^2 }
\end{eqnarray}
\end{mathletters}
We are interested in low energy phenomena; this implies that $k$ and $k+q+P$
are near 0 or $P$.  In this case we may approximate:
\begin{equation}
  \cosh(\theta_k) = (\mu/2\omega_k)^{1/2} (1+\omega_k/2\mu)
\end{equation}
\begin{equation}
  \sinh(\theta_k)=sgn(\gamma_k)(\mu/2\omega_k)^{1/2} (1-\omega_k/2\mu)
\end{equation}
Then near $q=0$ we have
\begin{eqnarray}
  \chi_q^{ss}(\omega) &=& \sum_k \frac{(\omega_k+\omega_{k+q})^2} {\omega_k
  \omega_{k+q}} \frac {b(\omega_k)-b(\omega_{k+q})}
  {\omega-\omega_k+\omega_{k+q}-i\epsilon} \cr &+& \sum_k \frac
  {(\omega_k-\omega_{k+q})^2} {\omega_k \omega_{k+q}}
  \frac{(1+b(\omega_k)+b(\omega_{k+q}))(\omega_k + \omega_{k+q})}
  {(\omega_k+\omega_{k+q})^2-(\omega+i\epsilon)^2}
\end{eqnarray}
\begin{eqnarray}
  \chi_q^{aa}(\omega) &=& \sum_k \frac{(\omega_k+\omega_{k+q+P})^2} {\omega_k
  \omega_{k+q+P}} \frac
  {b(\omega_k)-b(\omega_{k+q+P})}{\omega-\omega_k+\omega_{k+q+P}-i\epsilon} \cr
  &+& \sum_k \frac {(\omega_k-\omega_{k+q+P})^2} {\omega_k \omega_{k+q+P}}
  \frac{(1+b(\omega_k)+b(\omega_{k+q+P}))(\omega_k + \omega_{k+q+P})}
  {(\omega_k+\omega_{k+q+P})^2-(\omega+i\epsilon)^2}
\end{eqnarray}
while near $q=P$
\begin{equation}
  \chi_q^{ss}(\omega) = 4 \sum_k \frac{\mu^2} {\omega_k \omega_{k+q}} \left[
  \frac {b(\omega_k)-b(\omega_{k+q})}{\omega-\omega_k+\omega_{k+q}-i\epsilon} +
  \frac{(1+b(\omega_k)+b(\omega_{k+q}))(\omega_k+\omega_{k+q})}
  {(\omega_k+\omega_{k+q})^2-(\omega+i\epsilon)^2} \right]
\end{equation}
\begin{equation}
  \chi_q^{aa}(\omega) = 4 \sum_k \frac{\mu^2} {\omega_k \omega_{k+q+P}} \left[
  \frac{b(\omega_k)-b(\omega_{k+q+P})}
  {\omega-\omega_k+\omega_{k+q+P}-i\epsilon} +
  \frac{(1+b(\omega_k)+b(\omega_{k+q+P}))(\omega_k + \omega_{k+q+P})}
  {(\omega_k+\omega_{k+q+P})^2-(\omega+i\epsilon)^2} \right]
\end{equation}
In all of these formulae there are important contributions from $k$ near 0 and
$k$ near $P$.

We now consider NMR.  To derive nuclear relaxation rates we take the standard
[\onlinecite{MMP}] hyperfine Hamiltonians describing how nuclei are coupled to
the electronic spins, write the spins in terms of bosons, and then compute the
appropriate boson correlation functions. We assume throughout that $T \ll
J_1$.

We begin with the yttrium. An yttrium nucleus sits halfway between two
nearest-neighbor CuO$_2$ planes and above the center of a plaquette formed by
four Cu atoms. We denote the hyperfine coupling to one spin by $D$.  Thus we
write the hyperfine Hamiltonian for yttrium,
\begin{equation}
  H_{hf}^Y = D \sum_{a, i=1..4} \vec S_i^{(a)}
\end{equation}
After performing the transformations of section II and Appendix A and
retaining only those terms capable of giving dissipation at NMR frequencies we
have
\begin{equation}
  H_{hf}^Y = D \sum_{{k_1}, {q}, \alpha, \beta} g(q) \cosh(\theta_{k+q} -
  \theta_{k}) ( s^{\dagger}_{{k+q}\alpha}\sigma_{\alpha\beta}
  a^{\phantom{\dagger}}_{{k+P}\beta} + h.c.  )
\end{equation}
with
\begin{equation}
  |g(q)| = 4 \cos(q_x/2) \cos(q_y/2)
\end{equation}
Here we have omitted an unimportant phase factor in $g$.  We now calculate the
relaxation rate in the usual way, from
\begin{equation}
  \lim_{\omega \rightarrow 0} \frac{1}{\omega}\int_0^\infty
  e^{i(\omega+i\epsilon)t}<[H^Y_{hf}(t), H^Y_{hf}(0)]>
\end{equation}
finding
\begin{equation}
  \frac{1}{{^Y T}_1 T} = \frac{2\pi D^2}{T} \sum_{{{k}} {q}} \frac {|g(q)|^2
  \cosh^2(\theta_{k}-\theta_{k+q}) \delta(\omega_{k} - \omega_{k+q})}
  {\sinh^2(\omega_{k}/2T)} = 2 \pi D^2 \lim_{\omega \rightarrow
  0}\frac{1}{\omega} \sum_q |g(q)|^2 \chi^{ss''}(q,\omega)
\label{eq:C-1/YT1T}
\end{equation}
We now consider the planar oxygen.  Each oxygen is located in a CuO$_2$ plane
and is in the center of a bond connecting two Cu sites. Thus
\begin{equation}
  H_{hf}^O = C \sum_{i=1,2} \vec S_i^{(a)}
\end{equation}
Expressing the spins in terms of bosons as was done for yttrium gives:
\begin{eqnarray}
  H_{hf}^O = \frac{C}{2} \sum_{{{k}},{q}} f(q) (
  &\cosh&(\theta_{{k}}+\theta_{k+q+P}) (
  s^{\dagger}_{{{k}}\alpha}\sigma_{\alpha\beta}
  s^{\phantom{\dagger}}_{{k+q+P}\beta} +
  a^{\dagger}_{{{k+P}}\alpha}\sigma_{\alpha\beta}
  a^{\phantom{\dagger}}_{{k+q}\beta} )\cr + &\cosh&(\theta_{k}-\theta_{k+q})
  s^{\dagger}_{{{k}}\alpha}\sigma_{\alpha\beta}
  a^{\phantom{\dagger}}_{{k+q+P}\beta} + h.c.)
\label{eq:C-OHyperfine}
\end{eqnarray}
with
\[
  |f(k)| = 2 \cos\left(\frac{q_x}{2}\right)
\]
Proceeding as we did with yttrium yields eq. (\ref{eq:1/OT1T}) of the text.
To evaluate this it is convenient to consider $q$ near 0 and $q$ near $P$
separately.  The case of $q$ near 0 goes through just as for yttrium, except
that the square of the form factor is 4, not 16, and one must add
$\chi^{aa''}$.  For $q$ near $P$ it is convenient to write $q=P+k_2-k_1$ and
to sum over $k_1$ and $k_2$. The integrals have contributions from $k_1,k_2$
near 0 and $P$.  The form factor becomes
\begin{equation}
  |f(k_1-k_2)|^2 = (k_1^x-k_2^x)^2=(k_1^2+k_2^2)/2
\end{equation}
where in the second equality we have done the angular integral and k stands
for either $k$ or $(k-P)$ as appropriate.  Now we have, from eqs.
(\ref{eq:A-BosonEnergy}) and (\ref{eq:A-Dispersion}),
\begin{equation}
  k^2=4(1-\gamma_k)=2(\omega_k^2-\omega_{+,-})/\mu^2
\end{equation}
where the gap is $\omega_+$ for k near 0 and $\omega_-$ for $k$ near $P$.
Putting this into eq. (\ref{eq:1/OT1T}) yields eq. (\ref{eq:oT1}).  Finally,
we consider the Cu relaxation rate. A Cu nuclear moment is believed to be
coupled to the spin on the same site, via a hyperfine coupling $A$, and to the
spins on the four nearest neighbor sites in the same plane, via a hyperfine
coupling $B$. Thus
\begin{equation}
  H^{Cu}_{hf} = A\vec S_0^{(1)} + B \sum_{i=1..4} \vec S_i^{(1)}
\end{equation}
After transforming to the boson representation we have
\begin{eqnarray}
  H_{hf}^{Cu} = \frac{1}{2} \sum_{{k} {q}} \left[ A - 4 B
  \gamma(q) \right] ( &\cosh&(\theta_{k}+\theta_{k+q+P}) (
  s^{\dagger}_{{k}\alpha}\sigma_{\alpha\beta}
  s^{\phantom{\dagger}}_{{k+q+P}\beta} +
  a^{\dagger}_{{k}\alpha}\sigma_{\alpha\beta}
  a^{\phantom{\dagger}}_{{k+q+P}\beta} )\cr +
  &\cosh&(\theta_{k}-\theta_{k+q})
  s^{\dagger}_{{k+q}\alpha}\sigma_{\alpha\beta}
  a^{\phantom{\dagger}}_{{k+q+P}\beta} + h.c.  )
\end{eqnarray}
Again we may construct the relaxation rate.  It is given by eq.
(\ref{eq:CuT1T}) and may be simply evaluated because the dominant contribution
is at $q$ near $P$.

\begin{figure}[h]
  \caption{Model system considered in this paper: two square arrays
  of spins with in-plane coupling $J_1$ and between-plane coupling
  $J_2$.}
  \label{fig:Lattice}
\end{figure}
\begin{figure}[h]
  \caption{$T=0$ Phase diagram of eq. (\protect{\ref{eq:Model}}) as described
  in Appendix B from the Schwinger boson mean field theory.
  }
\end{figure}
\begin{figure}[h]
  \caption{
  Temperature dependence (in units of $J_2$) of the
  Schwinger boson gaps $\omega_+$ and $\omega_-$ defined in eqs.
  (\protect{\ref{eq:Defomega+}}) and (\protect{\ref{eq:Defomega-}}) and
  computed as described in Appendix B, for parameters such that at $T=0$ the
  model is at the phase boundary for small $J_2$ and $S_c - S$.  The
  temperature $T^* \sim J_2$ separates the low-$T$ coupled planes regime from
  a high-$T$ decoupled planes regime; in the mean field theory there is a
  second order phase transition at $T^*$; we believe fluctuations would
  convert this to a smooth crossover.
  }
\end{figure}
\begin{figure}
  \caption{
  Temperature dependence of uniform susceptibility and copper, oxygen and
  yttrium relaxation rates calculated from mean field theory for parameters
  used in constructing fig. 3.  The temperature $T^*$ at which the model
  crosses over from the decoupled-planes to the coupled-planes critical
  regimes is indicated.
  }
\end{figure}
\begin{figure}
  \caption{
  Frequency dependence of antisymmetric and symmetric susceptibilities
  calculated from mean field theory for several different temperatures.
  }
\end{figure}
\begin{figure}
  \caption{
  Different regimes of behavior of scaling theory and temperature dependence
  of relaxation rates in each regime.
  }
\end{figure}


\begin{thebibliography}{10}

\bibitem[(c)]{ETH} Address from September 1st 1993 on: Theoretische
Physik, ETH H\"onggerberg, CH-8093 Z\"urich, Switzerland

\bibitem{structure} T. Siegrist, S. Sunshine, D. W. Murphy, R. J. Cava and
S. M. Zahurak, Phys. Rev {\bf B35}, 7137 (1989).

\bibitem{MMP}
A. J. Millis, H. Monien and D. Pines, Phys. Rev. {\bf B42}, 996 (1991),
H.~Monien, D.~Pines and M.~Takigawa, Phys. Rev. {\bf 43}, 258 (1991)

\bibitem{Rossad}
J. Rossad-Mignod, L. P. Regnault, C. Vettier, P. Bourges, P. Burlet, J. Bossy,
J. Y. Henry and G. Lapertot, Physica {\bf C185-189}, 86 (1991).

\bibitem{Tranquada} J. M. Tranquada, P. M. Gehring, G. Shirane, S. Shamoto and
M. Sato, Phys. Rev. {\bf B46}, 5561, (1992).

\bibitem{Mook}
H. A. Mook, M. Yethiraj, G. Aeppli and T. Mason, Phys. Rev. Lett. {\bf 70},
3490 (1993).

\bibitem{Takigawa}
M. Takigawa et. al. Phys. Rev {\bf B42}, 243 (1991).

\bibitem{MM PRL}
A. J. Millis and H. Monien, Phys. Rev. Lett. {\bf 70}, 2810 (1993)

\bibitem{spingap}
T. M. Rice in the Proceedings of the ISSP Symposium on the Physics and
Chemistry of Oxide Superconductors, Tokyo (1991), Springer Verlag (1991).

\bibitem{Sokol}
A. Sokol and D. Pines, Phys. Rev. Lett. {\bf 71} 2813, (1993), and V. Barzykin,
D. Pines, A. V. Sokol and D. Thelen, unpublished.

\bibitem{CHN}
S.Chakravarty, B. I. Halperin and D. R. Nelson, Phys. Rev. Lett. {\bf 60},
1057 (1988) and S. Chakravarty, B. I. Halperin and D. R. Nelson, Phys. Rev
{\bf B39}, 7443 (1988).

\bibitem{Sachdev}
S. Sachdev and J. Ye, Phys. Rev. Lett. {\bf 69}, 2411 (1992)
and A. V. Chubukov and S. Sachdev Phys. Rev. Lett. {\bf 71}, 169 (1993).

\bibitem{LRO}
N. W. Preyer, R. J. Birgeneau, C. Y. Chen, D.  Gabbe, H. P. Jensen, M. A.
Kastner, P. J. Picone and T. Thio, Phys. Rev. {\bf B42}, 11563 (1989).

\bibitem{Hertz}
J. A. Hertz, Phys. Rev {\bf B14}, 1165, (1976).

\bibitem{AJM}
A. J. Millis, Phys. Rev {\bf B}48 7183 (1993).

\bibitem{Arovas Auerbach}
D. Arovas and A Auerbach, Phys. Rev. {\bf B38}, 316 (1988).

\bibitem{Sachdev and Read}
S. Sachdev and N. Read, Int. J. Mod. Phys. {\bf B5}, 219 (1991).

\bibitem{Hida}
K. Hida, J. Phys. Soc. Jpn. {\bf 61}, 1013 (1992).

\bibitem{Sandvik}
A. Sandvik, private communication

\bibitem{Chubukov}
A. V. Chubukov, Phys. Rev. {\bf B44}, 12318 (1991).

\bibitem{Chakravarty}
S. Chakravarty, in {\bf High Temperature Superconductivity: Proceedings},
K.~Bedell, D.~Coffey, D.~E. Meltzer, D.~Pines and J.~R.~Schriefffer, eds.
(Addison Wesley: Redwood City, CA), 179 (1990) and S.~Chakravarty and
R.~Orbach, Phys. Rev. Lett. {\bf 64}, 224 (1990)

\bibitem{eta}
P. Peczak, A. M. Ferrenberg and D. P. Landau, Phys. Rev. {\bf B43}, 6087
(1991).

\bibitem{Forster}
D. Forster, D. R. Nelson and M. Stephen, Phys. Rev. {\bf A16}, 732 (1977).

\bibitem{Pennington}
C. H. Pennington and C. P. Slichter, Phys. Rev. Lett. {\bf 66}, 381 (1991).

\bibitem{Machi}
T. Machi I. Tomeno, T. Miyatake, N. Koshizuka, S. Tanaka, T. Imai and H.
Yasuoka, Physica {\bf C173}, 32 (1991).

\bibitem{Pickett}
W. E. Pickett, Rev. Mod. Phys. {\bf 61}, 433 (1991), see especially section
IV-D and fig. 30.

\bibitem{Argonne}
R. Liu, B. W. Veal, A. P. Paulikas, J. W. Downey, P. J. Kostic, S. Fleshler,
U. Welp, C. G. Olson, X. Wu, A. J. Arko and J. Joyce, Phys. Rev {\bf B46},
11056 (1992).

\bibitem{Takigawa3}
For $T_2$, M. Takigawa et. al., unpublished; for $T_1$ see ref.
[\onlinecite{Takigawa}].

\bibitem{Imai}
T. Imai, C. P. Slichter, K. Yoshimura, M. Katoh and K. Kosuge, Phys. Rev. Lett.
{\bf 71}, 1254 (1993).

\bibitem{Singh}
R. Glenister, R. R. P. Singh and A. Sokol, unpublished.

\bibitem{Aeppli}
S.~Hayden, G.~Aeppli, R.~Osborn, A.~D.~Taylor, T.~G.~Perring, S.~W.~Cheong and
Z.~Fisk, Phys. Rev. Lett. {\bf 67}, 3622 (1991).

\bibitem{Imai2}
T. Imai, C. P. Slichter, K. Yoshimura and K. Kosuge, Phys. Rev. Lett. {\bf 70},
1002 (1993).

\bibitem{Forsterbook}
D. Forster {\bf Hydrodynamic Fluctuations, Broken Symmetry and
Correlation Functions}, (Benjamin Cummings:  Reading, PA) 1975.

\bibitem{J}
There is some disagreement in the literature about the value of J appropriate
to $YBa_2Cu_3O_{6.0}$.  The estimate $.1eV < J < .12eV$ is given in sec 2.3 of
a recent review (D. C. Johnston, J. M. M. M. 100, p. 218 (1991)).  The J value
for $La_2CuO_4$ is
0.13eV.

\bibitem{critical}
A. J. Millis and H. Monien, Phys. Rev. {\bf B45}, 3059 (1992).  Hyperfine
couplings are given in a convenient dimensionless form in Table 1 of this work,
and the observed ratio of Cu to O relaxation rates in $YBa_2Cu_3O_{6.6}$ at
$T=300K$ is quoted to be
16; we assume the ratio is similar in $YBa_2Cu_4O_8$.   The quantity defined
here as $A_{AF}$ is related to B in Table 1 of the cited work via
$A_{AF}=4B(1-\alpha )/(1+ \alpha )$. The bounds $0.2 < \alpha < 0.3$  were
obtained from data in this work also.
  The quantity defined here as $C_{AF}$ is actually a tensor with different
principal axes along and perpendicular to the Cu-O bond. For the estimates used
here we set $C_{AF} =C$ and took the value for fields along the Cu-O bond (as
appropriate for relax
ation with fields in the c-direction).  The ratio $A_{AF}/C_{AF}$ then turns
out to be 4 so at $T=300K$ where the predicted $\xi =4$ the predicted ratio is
about 256.

\bibitem{comment}
S. Sachdev and A. V. Chubukov, Phys. Rev. Lett. in press (1993).

\bibitem{SS}
B. I. Shraiman and E. D. Siggia, Phys. Rev. {\bf B42} 2485 (1990).

\bibitem{Brinkman}
H. Zimmermann, M. Mali, M. Bankay and D. Brinkmann, Physica C {\bf 185-189},
1145 (1991).  The relevant data are in fig. 3.  To convert Knight shifts to
susceptibilities we used the hyperfine couplings given in ref.
[\onlinecite{critical}] for YBa$_2$Cu$_3$O$_7$ with the quantity
$\chi_0/\mu_B^2$ of ref [\onlinecite{critical}] taken to be $2.7 states/eV-Cu$.


\bibitem{Takigawa2}
M. Takigawa, J. L. Smith and W. Hults, unpublished.

\end{thebibliography}
\end{document}